\documentclass[fleqn,usenatbib]{mnras}

\usepackage[T1]{fontenc}
\usepackage{ae,aecompl}

\usepackage{graphicx}	% Including figure files
\usepackage{amsmath}	% Advanced maths commands
\usepackage{amssymb}	% Extra maths symbols
\usepackage{times}	
\usepackage{makecell}
\usepackage{multirow}

\newcommand{\cf}{cf.~}
\newcommand{\ie}{i.e.,~}
\newcommand{\eg}{e.g.,~}

\title[kHz QPOs as oscillation tori modes: I.]{Kilohertz QPOs in low-mass
  X-ray binaries as oscillation modes of tori around neutron stars: I.}

\author[Marcio de Avellar et al.]{Marcio G B de
  Avellar$^{1,2}$,\thanks{E-mail: mgb.avellar@iag.usp.br}
Oliver Porth$^{2}$,
Ziri Younsi$^{2}$,
Luciano Rezzolla$^{2,3}$
\\
$^{1}$Instituto de Astronomia, Geof\'isica e Ci\^encias Atmosf\'ericas,
Universidade de S\~ao Paulo, 05508-090, S\~ao Paulo, Brasil\\
$^{2}$Institut f\"ur Theoretische Physik, Johann Wolfgang
Goethe-Universit\"at, Max-von-Laue-Str. 1 60438, Frankfurt am Main,
Germany\\
$^{3}$Frankfurt Institute for Advanced Studies,
  Ruth-Moufang-Strasse 1, 60438 Frankfurt am Main, Germany
}

\date{\today}
\pubyear{2017}

\begin{document}
\label{firstpage}
\pagerange{\pageref{firstpage}--\pageref{lastpage}}
\maketitle

\begin{abstract}
There have been many efforts to explain the dynamical mechanisms behind
the phenomenology of quasi-periodic oscillations (QPOs) seen in the X-ray
light curves of low-mass X-ray binaries. Up to now, none of the models
can successfully explain all the frequencies observed in the power
spectral density of the light curves. After performing several
general-relativistic hydrodynamic simulations of non-selfgravitating
axisymmetric thick tori with constant specific angular momentum
oscillating around a neutrons star such as the one associated with the
low-mass X-ray binary 4U 1636-53, we find that the oscillation modes give
rise to QPOs similar to those seen in the observational data. In
particular, when matching pairs of kilohertz QPOs from the numerical
simulations with those observed, certain combinations reproduce well the
observations, provided we take a mass for the neutron star that is
smaller than what generally assumed. At the same time, we find that tori
with constant specific angular momentum cannot match the entire range of
frequencies observed for 4U 1636-53 due to physical constraints set on
their size. Finally, we show that our results are consistent with the
observed shifts in QPO frequency that could accompany state transitions
of the accretion disc.
\end{abstract}


%%%%%%%%%%%%%%%%%%%%%%%%%%%%%%%%%%%%%%%%%%%%%%%%%%

%%%%%%%%%%%%%%%%% BODY OF PAPER %%%%%%%%%%%%%%%%%%

\section{Introduction}
\label{intro}

Several X-ray variability components can be seen in the power spectral
density (PSD) of the X-ray light curves of neutron-star and black-hole
low-mass X-ray binaries (NS- and BH-LMXBs, respectively). Among other
components, there is a vast set of quasi-periodic oscillations (QPOs)
ranging from $\sim 0.01$ to $\sim 1000$ Hz, whose physical origin
(origins) is still an open problem \citep[see][for a comprehensive review
  of QPOs in LMXBs]{vanDerKlis2006}. Overall, there are two broad
classes of models that address the mechanism behind the QPOs with
different degrees of success. The first one relies on orbital motions
(including the general-relativistic epicyclic motions) of matter around
the compact object and their timescales \citep[see, for
  example,][]{Miller1998,Stella1999,Abramowicz2004b}. The second
class, the focus of this study, interprets the QPOs as oscillations
and/or flow instabilities of some kind of an accretion disc orbiting
around the central object \citep[see, for
  example,][]{Rezzolla_qpo_03a}. The study of accretion flows around
compact objects dates back to the 1970s with the pioneering works of
\citet{Shakura1973} and \citet{Novikov:1973} when considering
geometrically thin accretion discs, and to the works of
\citet{Fishbone76,Abramowicz78,Kozlowski1978} in the opposite case of
geometrically thick discs. The latter are possible solutions of a
stationary and axisymmetric fluid that undergoes a non-Keplerian circular
motion around a compact object. Despite the variety of mechanisms
proposed so far, the scarcity of the observational data and the weak
constraints it poses, do not seem to favour one model over the other.

Astronomical observations over the years have revealed that the
frequencies of different QPOs observed in a given source correlate with
each other through very specific patterns and, more importantly, that the
same patterns are then seen across systems of the same type
\citep[e.g.,][]{vanStraaten2002, vanStraaten2003, Reig2004,
  vanStraaten2005, Altamirano2008}.  The correlations seen in
the PSDs of different types of X-ray binaries, i.e., containing black
holes, neutron stars or white dwarfs, seem to indicate that a common
component of these systems, most likely the accretion disc, is
responsible for the origin of these variability features
\citep[e.g.,][]{Wijnands1999, Psaltis1999, Mauche2002,
  Warner2002}. Within this accretion-disc interpretation, the highest
frequency QPOs, known as the lower ($\nu_{l}$) and upper ($\nu_{u}$)
kilohertz QPOs (kHz QPOs), are thought to reflect the properties of the
accretion flow in the vicinity of black holes and neutron stars, where
effects of the general relativity are expected
\citep[see][]{vanDerKlis2006}.

Other properties of the QPOs, besides the frequency itself, like the
fractional amplitude, quality factor, and time/phase lags, also depend on
the spectral state of the source, which can be parametrized by the
position of the source in what is called the ``colour-colour diagram''
(CCD) in X-rays \citep[e.g.,][]{Hasinger1989, Wijnands1997,
  vanStraaten2002, diSalvo2003, Mendez2006, Altamirano2008,
  deAvellar2016}. The changes in the the position of the source
within the CCD are therefore expected to reflect, at least qualitatively,
different configurations of the flow near the central compact object
\citep[see][for example]{Altamirano2008}.

From the theoretical point of view, geometrically thick discs achieved
astrophysical relevance once it was demonstrated that external
perturbations cause them to oscillate periodically, opening new avenues
for the interpretation of the phenomenology of QPOs \citep[e.g.,][and
  references therein for the vast literature that has developed over the
  years]{Zanotti03, Rezzolla_qpo_03a, Lee2004, Abramowicz2004b,
  Zanotti05, Blaes2006, Remillard2006, Rezzolla_qpo_03a, Abramowicz2003,
  Bursa2004, Kluzniak2004, Schnittman06, Blaes2007,
  Montero2012, Mazur2013, Vincent2014, Bakala2015,
  Mishra2017}. In particular, \citet{Rezzolla_qpo_03b} identified
the frequencies in the X-ray spectra of BH-LMXBs as inertial-acoustic
modes, or $p$-modes, of a relativistic geometrically-thick toroidal
accretion disc. The model is successful in reproducing approximately the
3:2 ratio seen in the range of the high-frequency (HF) QPOs observed in
these kinds of systems. Further investigations of the properties of
axisymmetric $p$-mode oscillations were later on performed by, for
example, \citet{Zanotti03,Zanotti05,Montero07,Montero2008,Montero2010},
through general-relativistic hydrodynamical (GRHD) simulations of both
non-selfgravitating and selfgravitating tori around compact objects.

This paper extends these simulations to NS-LMXBs, with particular
application to the Atoll\footnote{For a description of the Z and Atoll
  sources, see \citet{Hasinger1989}} source 4U 1636--53, for which more
than 15 years of X-ray data have been collected and analysed in their
many aspects \citep[see, for example,][for an incomplete set of
  references; see references therein]{diSalvo2003, Casares2006,
  Barret2006, Altamirano2008, Lin2011, Sanna2012,
  Artigue2013, deAvellar2013, Sanna2014, Lyu2014,
  deAvellar2016, Zhang2017, Ludlam2017}. More
specifically, this paper details how we build a sequence of 21
relativistic and axisymmetric tori with constant specific angular
momentum and follow their oscillations triggered by a small radial and
vertical velocity perturbation.

The plan of the paper is as follows. In Sec. \ref{method}, we discuss how
frequencies are identified from the simulation data and cross-compared to
the observational $\nu_{l}$-$\nu_{u}$ relation (Section \ref{res}). The
results are discussed in Sec. \ref{discuss} elucidating the implications
of our analysis with application to the changing properties of the kHz
QPOs as the source moves across the CCD. Throughout we adopt the
geometrized system of units $G=c=1$, with $G$ Newton's constant and $c$
the speed of light.

\section{Numerical setup}
\label{numSetup}

\subsection{Initial configurations of the tori}
\label{initConfigTori}

As a first step we describe the construction of the initial
equilibrium torus solutions of which the oscillation frequencies are
analysed in the remainder of this paper.  We consider tori as
stationary non-selfgravitating fluid configurations that can be built
around a compact object as a solution for the GRHD equations \citep[as
  in][for example]{Kozlowski1978}. To simulate the dynamics of this system we
solve the usual GRHD equations written in their covariant form
\begin{eqnarray}
\label{eq:current_cons}
\nabla _{\mu}\big(\rho u^{\mu}\big) = 0 \,, \\
\label{eq:em_cons}
\nabla _{\mu}T^{\mu\nu} = 0 \,,
\end{eqnarray}
where $\rho$ is the rest-mass density, $u^{\mu}$ are the (contravariant)
components of the fluid four-velocity and $T^{\mu\nu}$ is the
energy-momentum tensor \citep[see, for example,][]{Rezzolla_book:2013}. 

To close the system of equations \eqref{eq:current_cons} and
\eqref{eq:em_cons} we assume an ideal-fluid equation of state
\citep{Rezzolla_book:2013} $p=\rho \epsilon({\hat{\gamma}}-1)$, where
$\epsilon$ is the specific internal energy and $\hat{\gamma}$ is the
adiabatic index.  The initial (test-fluid) equilibrium is obtained as isentropic
configuration, thus we set the entropy $\kappa=p/\rho^{\hat{\gamma}}$
to an arbitrary constant (we have taken $\kappa=0.001$)
 \footnote{We recall that ideal-fluid and the
    polytropic equations of state coincide for isentropic
    transformations as it is the case for the initial data
    \cite{Rezzolla_book:2013}.}.  As a result, the specific enthalpy
  $h(\rho,p)$ is expressed as
\begin{equation}
h = 1 + \epsilon + \frac{p}{\rho} = 1+
\frac{\hat{\gamma}}{\hat{\gamma} - 1}\frac{p}{\rho}\,.
\end{equation} 
Since the fluid is following a non-Keplerian motion, the distribution of
specific angular momentum $\ell := u_{\phi}/u_{t}$ needs to be specified,
where $u_{\phi}, u_{t}$ are the azimuthal and time components of the
four-velocity, respectively. The simplest choice in this case, and the
one that has been historically best studied, is to set $\ell = {\rm
  const.}$, so that the maximum equatorial torus size for the marginally
stable non-accreting configuration is determined by the value of the
specific angular momentum distribution for the fluid
\citep{Font02a,Daigne04,Rezzolla_book:2013}.

%%%%%%%%%%%%%%%%%%%%%%%%%%%%%%%%%%%%%%
%
% Table 1
%
%
\begin{table*}
\caption{Parameters of the models built and simulated in this
  work. Reported are: the resolution, specific angular momentum
  ($\ell_{o}$), the position of the cusp ($r_{\mathrm{cusp}}$), the
  position of the inner edge of the torus ($r_{\mathrm{in}}$), the
  position of the centre of the torus ($r_{\mathrm{centre}}$), the outer
  edge of the torus ($r_{\mathrm{out}}$) and the size of the torus
  ($\ast$ $\ast$); for each quantity the corresponding units are
  indicated in the square. Note that only the odd-numbered models from
  \texttt{Tor.03} to \texttt{Tor.21} have been simulated numerically.}
\begin{tabular}{crccccrr}
\hline & & & & & & & \\ Models & resolution $~$ $~$ & $\ell_{o}$
       [$M^{2}$] & $r_{\mathrm{cusp}}$ [M] & $r_{\mathrm{in}}$ [$M$] &
       $r_{\mathrm{centre}}$ [$M$] & $r_{\mathrm{out}}$ [$M$] & size [km]
       ($\ast$ $\ast$) \\[2pt] \hline & & & & & & & \\
%    Mod 01 & 5248 $\times$ 3520 $\ast$ & 3.6800 & 5.6321 & 5.7419 & 6.4098 & 6.8850 & 2.87 $~$$~$$~$$~$$~$ \\[2pt]
%    Mod 02 & 3392 $\times$ 2240 $\ast$  & 3.6866 & 5.4742 & 5.6357 & 6.6157 & 7.3942 & 4.41 $~$$~$$~$$~$$~$ \\[2pt]
    \texttt{Tor.03} & 2624 $\times$ 1728 $\ast$  & 3.6932 & 5.3607 & 5.5617 & 6.7774 & 7.8297 & 5.69 $~$$~$$~$$~$$~$ \\[2pt]
%    \texttt{Tor.04 & 2176 $\times$ 1472 $\ast$  & 3.6998 & 5.2689 & 5.5034 & 6.9173 & 8.2349 & 6.86 $~$$~$$~$$~$$~$ \\[2pt]
    \texttt{Tor.05} & 1920 $\times$ 1280 $\ast$  & 3.7064 & 5.1908 & 5.4529 & 7.0437 & 8.6271 & 7.97 $~$$~$$~$$~$$~$ \\[2pt]
%    \texttt{Tor.06} & 1664 $\times$ 1088 $\ast$  & 3.7130 & 5.1221 & 5.4090 & 7.1608 & 9.0138 & 9.05 $~$$~$$~$$~$$~$ \\[2pt]
    \texttt{Tor.07} & 1472 $\times$ 960 $\ast$  & 3.7196 & 5.0606 & 5.3693 & 7.2707 & 9.4006 & 10.12 $~$$~$$~$$~$$~$ \\[2pt]
%    \texttt{Tor.08 & 1344 $\times$ 896 $\ast$  & 3.7262 & 5.0047 & 5.3300 & 7.3752 & 9.7938 & 11.21 $~$$~$$~$$~$$~$ \\[2pt]
    \texttt{Tor.09} & 1216 $\times$ 832 $\ast$  & 3.7328 & 4.9533 & 5.2951 & 7.4753 & 10.1923 & 12.29 $~$$~$$~$$~$$~$ \\[2pt]
%    \texttt{Tor.10} & 1088 $\times$ 704 $\ast$  & 3.7394 & 4.9056 & 5.2637 & 7.5716 & 10.5982 & 13.39 $~$$~$$~$$~$$~$ \\[2pt]
    \texttt{Tor.11} & 1040 $\times$ 700 $~$ & 3.7460 & 4.8612 & 5.2306 & 7.6649 & 11.0196 & 14.53 $~$$~$$~$$~$$~$ \\[2pt]
%    \texttt{Tor.12 & 960 $\times$ 640 $~$ & 3.7526 & 4.8194 & 5.2002 & 7.7556 & 11.4532 & 15.70 $~$$~$$~$$~$$~$ \\[2pt]
    \texttt{Tor.13} & 900 $\times$ 600 $~$ & 3.7592 & 4.7801 & 5.1721 & 7.8439 & 11.9001 & 16.89 $~$$~$$~$$~$$~$ \\[2pt]
%    \texttt{Tor.14 & 840 $\times$ 560 $~$ & 3.7658 & 4.7428 & 5.1460 & 7.9302 & 12.3638 & 18.12 $~$$~$$~$$~$$~$ \\[2pt]
    \texttt{Tor.15} & 780 $\times$ 520 $~$ & 3.7724 & 4.7075 & 5.1170 & 8.0148 & 12.8531 & 19.42 $~$$~$$~$$~$$~$ \\[2pt]
%    \texttt{Tor.16 & 720 $\times$ 480 $~$ & 3.7790 & 4.6738 & 5.0897 & 8.0977 & 13.3631 & 20.77 $~$$~$$~$$~$$~$ \\[2pt]
    \texttt{Tor.17} & 680 $\times$ 460 $~$ & 3.7856 & 4.6416 & 5.0575 & 8.1793 & 13.9106 & 22.22 $~$$~$$~$$~$$~$ \\[2pt]
%    \texttt{Tor.18 & 640 $\times$ 420 $~$ & 3.7922 & 4.6108 & 5.0304 & 8.2595 & 14.4770 & 23.71 $~$$~$$~$$~$$~$ \\[2pt]
    \texttt{Tor.19} & 600 $\times$ 400 $~$ & 3.7988 & 4.5812 & 5.0073 & 8.3386 & 15.0653 & 25.25 $~$$~$$~$$~$$~$ \\[2pt]
%    \texttt{Tor.20 & 560 $\times$ 380 $~$ & 3.8054 & 4.5528 & 4.9808 & 8.4166 & 15.6977 & 26.90 $~$$~$$~$$~$$~$ \\[2pt]
    \texttt{Tor.21} & 560 $\times$ 380 $~$ & 3.8062 & 4.5494 & 4.9771 & 8.4262 & 15.7797 & 27.12 $~$$~$$~$$~$$~$ \\[2pt]
\hline
\hline
\multicolumn{8}{l}{($\ast$) resolution reached in the torus area after three levels of mesh refinement.} \\
\multicolumn{8}{l}{($\ast$ $\ast$) assuming a neutron star mass of $1.7~M_{\odot}$.}
\end{tabular}
\label{tab:tori_pars}
\end{table*}
%%%%%%%%%%%%%%%%%%%%%%%%%%%%%%%%%%%%%%

As mentioned in Sec. \ref{intro}, we use the NS-LMXB in 4U 1636--53 as
our study case, and the torus should therefore not penetrate the surface
of the star, assumed to have a mass which falls in the uncertainty
bracket suggested by \citet{Casares2006}, \ie $M =
1.7\,M_{\odot}$. Assuming the representative nuclear-physics equation of
state SLy4 of \citep{Douchin01}, to such a mass would correspond to a
neutron star with radius of $11.42$ km when nonrotating.

Since the exterior spacetime is given by the Schwarzschild solution with
$M$ the mass of the neutron star, we choose values of the constant
specific angular momentum $\ell \in [3.68, 3.81]$, noting that a torus
with $\ell = 3\sqrt{6}/2 \simeq 3.67$ is actually a ring
\citep{Rezzolla_book:2013}. We recall, in fact, that the position of the
cusp and the centre, which are the location where the specific angular
momentum coincides with the corresponding Keplerian one, are both at
$6\,M$ for a nonrotating central star of mass $M$ and the orbital
frequency of the fluid in this case tends to the orbital frequency of a
test particle. Hence, a torus with $\ell\geq 3.81$ would penetrate the
stellar surface, \ie it would have $r_{\mathrm{cusp}} <
r_{\mathrm{star}}$.

Also, in order to minimise mass loss at each oscillation, we employ a
jump in the effective potential, $\Delta W := W_{\mathrm{in}}
-W_{\mathrm{cusp}}$, where $W_{\mathrm{in}}$ is the potential at the
inner edge of the torus and $W_{\mathrm{cusp}}$ is the potential at the
cusp of the of the potential well, specifying $W_{\mathrm{in}}$ in a way
that the size of the torus is $\simeq 90\%$ of the maximum equatorial
torus size for the marginally stable non-accreting configuration. Hence,
the inner edge of each torus is always safely below the inner edge of the
marginally stable one. Following the prescription described above and
which resembles what suggested by \citet{Font02a}, we have built 21
axisymmetric torus models of increasing size; the main properties of some
of these models are summarised in Table \ref{tab:tori_pars}.

The oscillations are triggered by specifying small radial and vertical
perturbations in the 4-velocity field given by
\begin{equation}
(\delta v^{i}) = \Bigg(0,\frac{\eta}{\sqrt{2\,g_{rr}}},
  \frac{\eta}{\sqrt{2\,g_{\theta\theta}}},0\Bigg) \,,
\end{equation} 
where $\eta := 0.01\, \sqrt{W_{\mathrm{in}} - W_{\mathrm{centre}}}$ and
$W_{\mathrm{centre}}$ is the potential at the centre of the torus and
which also corresponds with the position of the maximum rest-mass
density, ensuring that the perturbation in the vertical and radial
directions have the same magnitude as can be readily seen by computing
the scalar product $\delta v^{i}\delta v_{i}= \eta^{2}$. Since the
quantity $\sqrt{W_{\mathrm{in}} - W_{\mathrm{centre}}}$ is proportional
to the ``escape velocity'' from the potential well, the perturbation in
the velocity is comparable across all torus models.

\subsection{Numerical approach}
\label{numAppr}

The simulations were carried out using the black-hole accretion code
\texttt{BHAC} \citep{Porth2017}, which solves the ideal
general-relativistic magnetohydrodynamic (GRMHD) equations in arbitrary
spacetimes, exploiting shock-capturing techniques and adaptive mesh
refinement. For a detailed description of the code and its capabilities
see \citet{Porth2017}. \texttt{BHAC} is a versatile code that allows the
user to freely choose the metric theory of gravity and coordinate
system. Here we use the Schwarzschild metric in modified Kerr-Schild
coordinates, whose main purpose is to remove the coordinate singularity
at the horizon but also to stretch the grid radially thus reducing the
computational cost.

Our grid extends from the neutron star radius at $r_{\mathrm{star}}
\simeq 4.549\,M$ to $30\,M$ in the radial direction and from
$\pi/2-0.628$ to $\pi/2+0.628$ in the $\theta$ direction. The resolution
of the grid was chosen in order resolve all tori with $\simeq 200$ cells
in their diameter. Since uniform resolution would have been impractical
for simulations of small tori, three levels of mesh refinement were
employed in the the smallest tori (\cf Table \ref{tab:tori_pars}).

\begin{figure*}
\centering
\includegraphics[width=0.95\columnwidth]{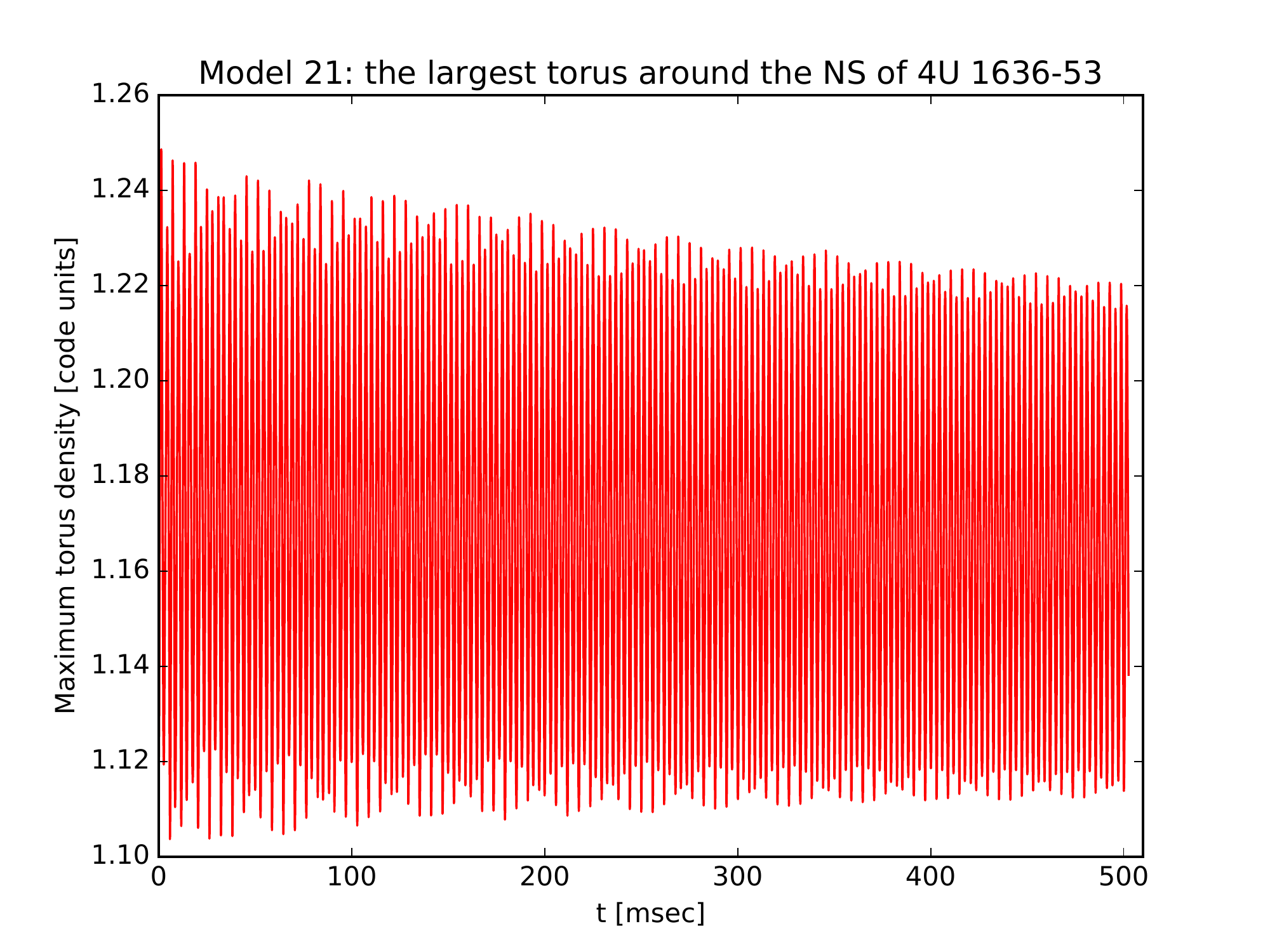}
\hskip 1.0cm
\includegraphics[width=0.95\columnwidth]{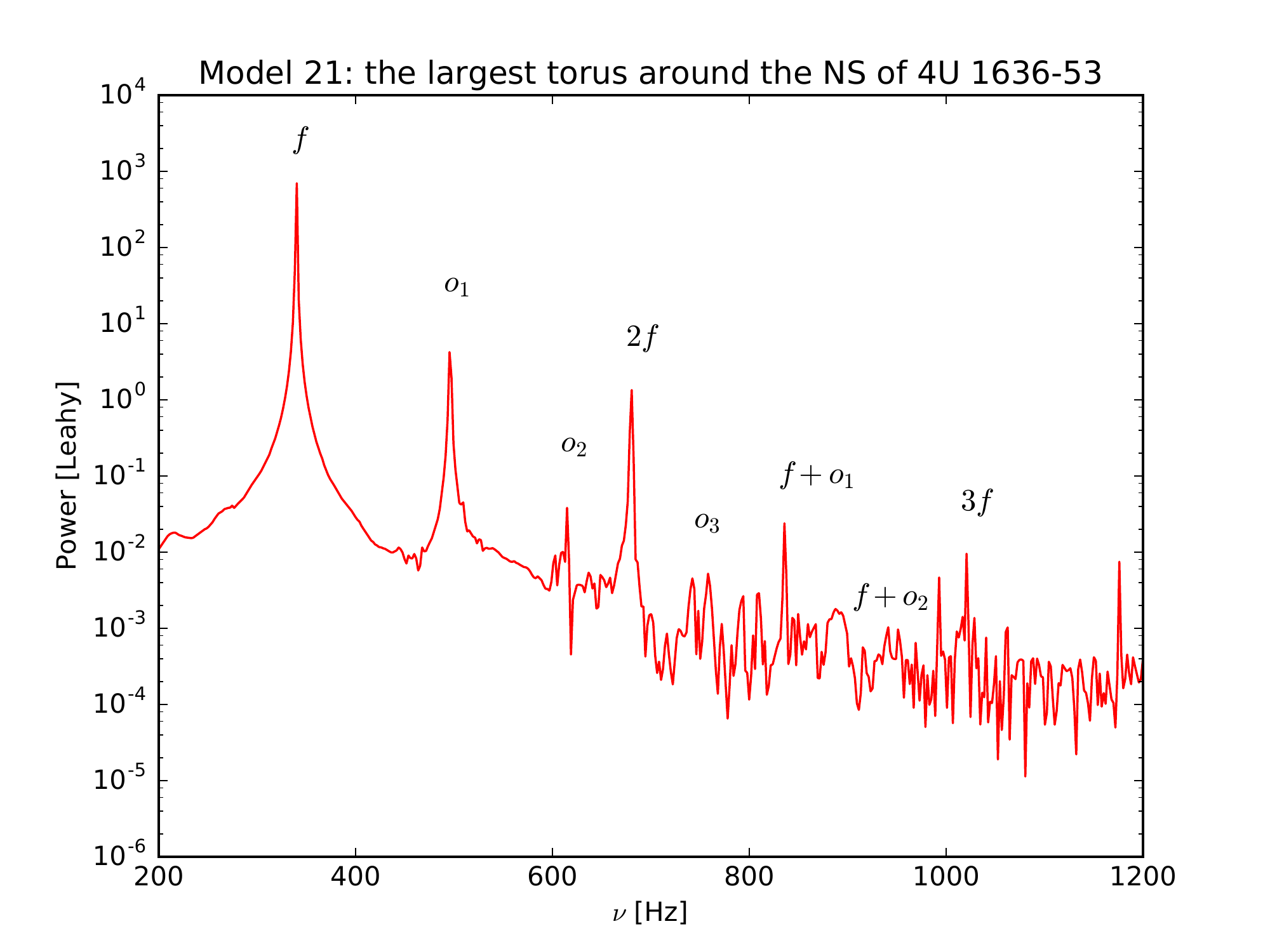}
\caption{Oscillations in the the maximum rest-mass density (left panel)
  and its corresponding PSD (right panel) of \texttt{Tor.21}, the largest
  torus around the neutron star in 4U 1636--53 (see Table
  \ref{tab:tori_pars}). The small damping in the density is due to $\sim
  1\%$ mass loss, which is also the largest measured in our
  simulations. Other tori show very similar behaviours, with the position
  and amplitude of the modes obviously depending on the specific torus
  considered.}
\label{fig:oscillationPSD}
\end{figure*}

\subsection{Adopted methodology for the data analysis}
\label{method}

Although we have constructed 21 models different initial models
representing stationary tori, our evolutions have been performed using
only the odd-numbered models from \texttt{Tor.03} to \texttt{Tor.21},
for a total of ten models simulated. This choice was dictated by the
need to reduce the computational costs but does not come at the cost
of reducing the generality of our results and conclusions. Also,
models smaller than \texttt{Tor.03} were not considered because they
turn out too small to be physically meaningful. We recall, in fact,
that as the torus' size goes to zero (both in the radial and in the
polar directions), the corresponding oscillations approach those of a
test particle at the position of the centre, which can be readily
computed analytically \citet{Rezzolla_qpo_03a, Zanotti03}.

We have therefore evolved each of the ten models up to $t = 60\,000\,M$,
which is equivalent to $\simeq 500$ orbital periods, and corresponds to a
timescale of $\sim 500\,{\rm ms}$ when assuming $M=1.7\,M_{\odot}$. Due
to the velocity perturbation, the tori will oscillate inside the
potential well, possibly leading to mass loss or induced accretion at
each period. Specifying $W_{\mathrm{in}}$ as mentioned in Sec.
\ref{initConfigTori} and as shown in Table \ref{tab:tori_pars} we
minimise this effect. As a result, the oscillation can survive in all
runs with only a minimal damping due to the mass loss; no significant
drift in frequencies is observed as a consequence of this prescription.

\citet{Rezzolla_qpo_03a} and \citet{Zanotti03} studied toroidal fluid
configurations around Schwarzschild black holes having a constant
specific angular momentum distribution; the analysis by
\citet{Rezzolla_qpo_03a} was semi-analytical and involved linear
perturbation theory, while the analysis by \citet{Zanotti03} was fully
numerical and involved general-relativistic hydrodynamic
simulations. Notwithstanding the two different approaches, both analyses
found that the $p$-modes of these configurations appear in frequencies
obeying the $2:3:4:...$ ratio. \citet{Zanotti05}, on the other hand,
studied through both linear analysis and simulations, toroidal fluid
configurations around Kerr black holes also having non-constant specific
angular-momentum distributions; as a result, a richer set of frequencies
was simulated and measured. The linear perturbative analysis formulated a
prediction about the eigenfrequencies of the $p$-modes and the ratio in
which they appear, thus providing a theoretical insight into where to
look for the frequencies coming from the simulations. In either case, the
authors studied the response of the $L_{2}$-norm of the density which as
a global quantity is particularly suitable for comparisons with the
results coming from the perturbative analysis.

In our analysis we have followed the periodic response of the central
rest-mass density of the tori ($\rho_{\mathrm{max}}$) to the velocity
perturbation. The choice for the central density is physically motivated
in the sense that it is where the concentration of matter is larger and
where most likely the variations in (X-ray) luminosity occur, i.e., where
the QPOs would come from \citep[see][]{Schnittman06, Vincent2014}.

We note that the responses of a given hydrodynamical quantity, \eg the
central rest-mass density and of its $L_{2}$-norm over the computational
domain $\Omega$, \ie $L_{2}(\rho) :=
\sqrt{{\int_{\Omega}\rho^{2}dV}/{\int_{\Omega}dV}}$, should be
equivalent. However, it was found that while the frequency of the peaks
remains the same for both quantities, the strength of the peaks and even
their appearance depend to varying degree on the quantity under
consideration. This fact raises the question of how to appropriately
identify the peaks to be considered in the analysis and as a consequence
a suitable strategy was developed for this purpose (see discussion in
Sec. \ref{sec:freq_iden} below). Figure \ref{fig:oscillationPSD} reports
the oscillations of the central rest-mass density $\rho_{\mathrm{max}}$
and its corresponding PSD for \texttt{Tor.21}, which the largest torus
around the neutron star in 4U 1636--53. The mass lost in this case is the
largest of all the models considered, but it is only of the order of
$1\%$.

\begin{figure}
\centering
\includegraphics[scale=0.40]{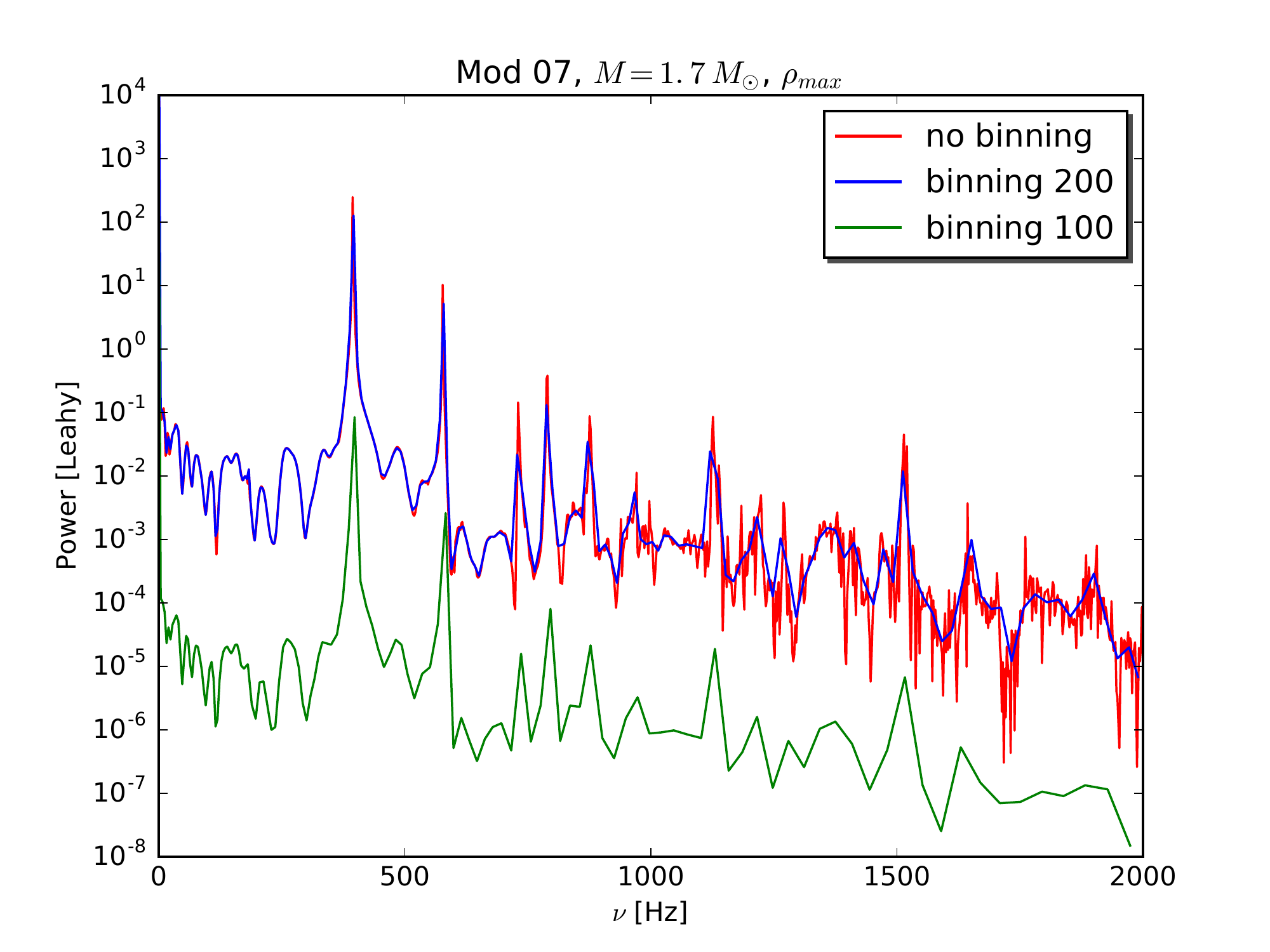}
\caption{Examples of binned and not binned PSDs of the maximum rest-mass
  density. The data refers to the model \texttt{Tor.21} and shows that
  even a low binning rate retains the main features of the PSDs.}
\label{fig:logBinning}
\end{figure}

\subsubsection{Frequency identification}
\label{sec:freq_iden}

In order to proceed with the identification of the modes observed, we
have followed the linear analysis by \citet{Rezzolla_qpo_03a,Zanotti03}
and \cite{Zanotti05}, and first identified and labelled the $p$-mode
fundamental frequency and its overtones using the same nomenclature, as
shown in the right panel of Fig. \ref{fig:oscillationPSD} for the
representative torus model \texttt{Tor.21}. Proceeding in this manner, we
were able to determine eight different frequencies of increasing value:
$f$, $o_{1}$, $o_{2}$, $2f$, $o_{3}$, $f+o_{1}$, $f+o_{2}$, and
$3f$. These peaks in the PSD correspond to the same frequencies
identified by \citet{Zanotti05} in their analysis, with the potential
addition of the third overtone labelled $o_{3}$, whose excess power is
however rather small.

\begin{figure*}
\centering
\includegraphics[width=0.95\columnwidth]{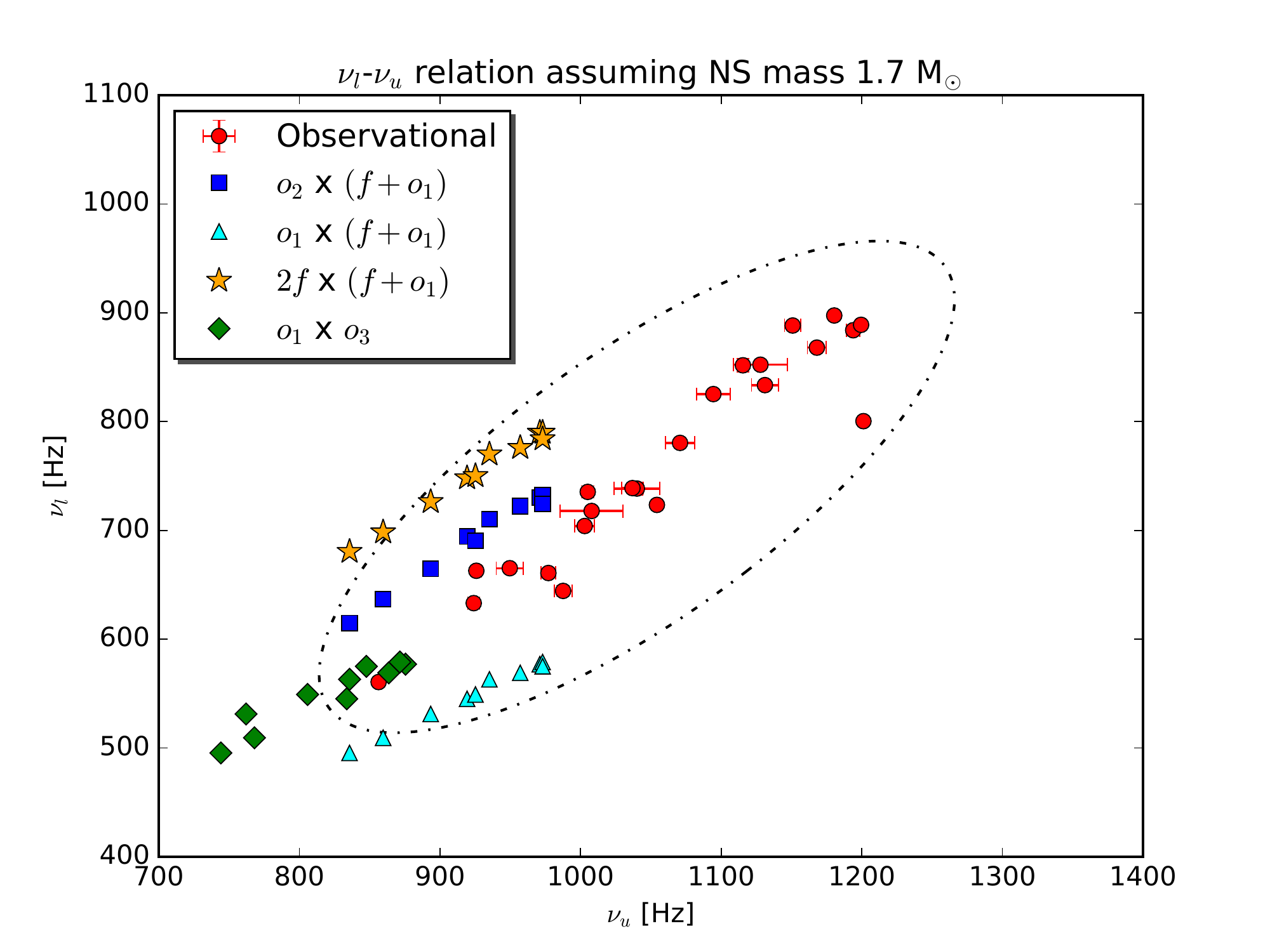}
\hskip 1.0cm
\includegraphics[width=0.95\columnwidth]{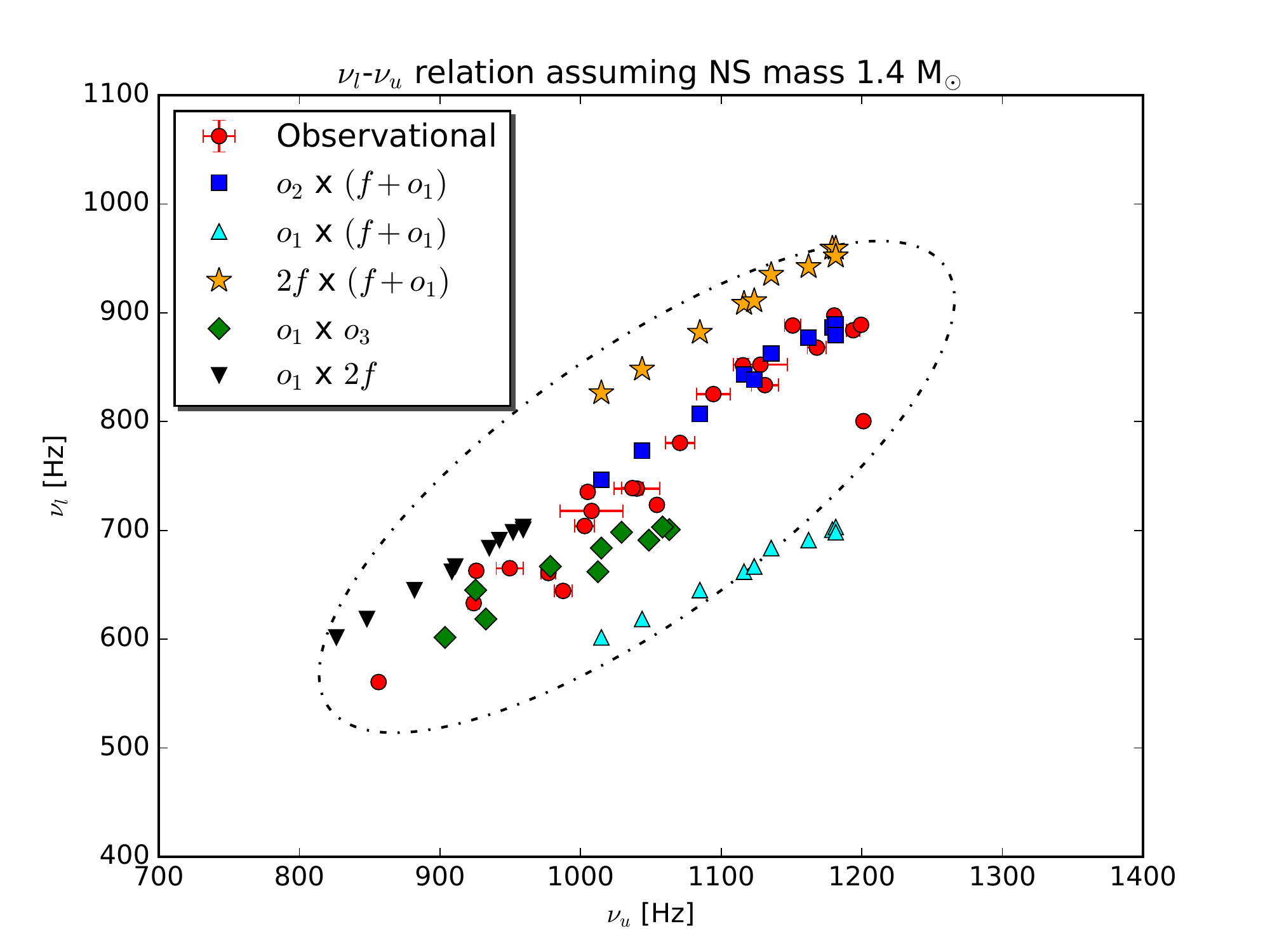}
\caption{Selected pairs of peaks in the PSD whose numerical relations are
  presented in Table \ref{tab:freqRatios}. A larger torus gives rise to
  lower characteristic frequencies, while higher masses for the central
  star result in lower frequencies. The lower the value of the neutron
  star mass, the higher the values of the frequencies of the peaks. The
  frequencies of the simulated peaks are scalable with the mass of the
  neutron star, $\nu_{1}/\nu_{2} = {M_{2}}/{M_{1}}$, as are the peak
  separations $\Delta\nu_{1}/\Delta\nu_{2} = {M_{2}}/ {M_{1}}$. The
  dot-dashed ellipse delineates a region around the observational
  relation where suitable choices of pairs of peaks are found.}
\label{fig:qpos_as_axisymmetric_oscillations}
\end{figure*}

However, we recall that not all frequencies are detected in all
models. Additional tests were performed to ensure that in this study of
the QPOs only reliable peaks were used. More information can be gained by
logarithmically binning our data and comparing the binned PSDs with the
non-binned ones\footnote{Rebinning by a factor $N$ means that $N$ points
  are taken and averaged into a single value}. Logarithmically rebinning
the data and averaging the same number of points per decade results in a
rebinning factor of $10^{1/N}$. Performing the binning correctly will add
power to a feature which is present in the data while smearing out the
noise. Hence, we have adopted two binning factors to determine whether a
given peak was genuine of the possible contamination from noise: $N=200$
(infrequent binning) and $N=100$ (frequent binning). An example of this
procedure is illustrated in Fig.  \ref{fig:logBinning} for
\texttt{Tor.07}.

In this way, a systematic study of the statistics of the various models
was performed through which we counted how often a given peak was
detected ``robustly'' through all simulation models. By ``robust'', we here
refer to a peak that needed to be detected for at least seven torus
models and both in the PSD of the maximum rest-mass density and in the
PSD of the corresponding $L_2$-norm. This validation was necessary to
obtain the confidence that the peak considered actually referred to a
physical mode of oscillation. Following these criteria for the
identification of the peaks in the PSD and for the determination of their
robustness, the following frequencies were ultimately selected: $f$,
$o_{1}$, $o_{2}$, $2f$, $o_{3}$ and $f+o_{1}$. The peaks associated with
the frequencies $f+o_{2}$ and $3f$ did not pass these tests and were
therefore not considered in the subsequent analysis, even though they
appear clearly in some models.

%%%%%%%%%%%%%%%%%%%%%%%%%%%%%%%%%%%%%%
%
% Table 2
%
%
\begin{table}
\caption{Ratio between the frequencies of the selected modes measured
  from the simulations. Each values reports the ratio between the
  frequency of the mode shown horizontally with the frequency of the mode
  shown vertically. Shown in boldface in the bottom right corner of the
  Table is the ratio of the frequencies of the upper and lower kHz QPOs,
  $\nu_{u}/\nu_{l}$, for the source NS-LMXB 4U 1636--53.}
\setlength\tabcolsep{3.30pt} \centering
\begin{tabular}{|c|c|ccccccc|}
\hline
 &            & \multicolumn{7}{c|}{$\nu_{u}$: higher kHz QPO}\\
\hline
 &            & $f$                      & $o_{1}$ & $o_{2}$   & $2f$ & $o_{3}$   & $f+o_{1}$ & ${\nu_{u}}$  \\
\hline
\multirow{7}{*}{\rotatebox{90}{$\nu_{l}$: lower kHz QPO}}   
 &            &                          &          &            &      &            &            &        \\
 & $f$        &                          &    1.46  & 1.81--1.86 & 2.00 & 2.10--2.23 & 2.46       &        \\    
 & $o_{1}$   &                          &          & 1.24--1.27 & 1.36 & 1.43--1.52 & 1.68       &        \\
 & $o_{2}$   &      		         &          &            & 1.08 & 1.15--1.21 & 1.32--1.36 &        \\
 & $2f$       &      		         &          &            &      & 1.05--1.11 & 1.23       &        \\
 & $o_{3}$   &     		         &          &            &      &            & 1.11--1.17 &        \\    
 & $\nu_{l}$ &     		         &          &            &      &            &            & \textbf{1.30--1.53} \\
\hline
\end{tabular}
\label{tab:freqRatios}
\end{table}
%%%%%%%%%%%%%%%%%%%%%%%%%%%%%%%%%%%%%%

\section{Results and discussion}
\label{res}

A first and possibly most important result of our analysis is that the
most prominent peaks of the oscillation modes of tori around neutron
stars obey specific ratios, as shown in Table \ref{tab:freqRatios}, and
in agreement with the linear analysis performed by
\citet{Rezzolla_qpo_03b} and by \citet{Zanotti03}. A summarising
representation of this conclusion is presented in Fig.
\ref{fig:qpos_as_axisymmetric_oscillations}, where we show the
observational $\nu_{l}\times\nu_{u}$ relation for the NS-LMXB 4U
1636--53, together with some choices of pairs of peaks (the frequencies
of the modes of oscillation) produced by the simulated oscillating
tori. In the following discussion we will concentrate on two specific
cases, as these are the most representative and interesting. 

\begin{figure}
\centering
\includegraphics[scale=0.40]{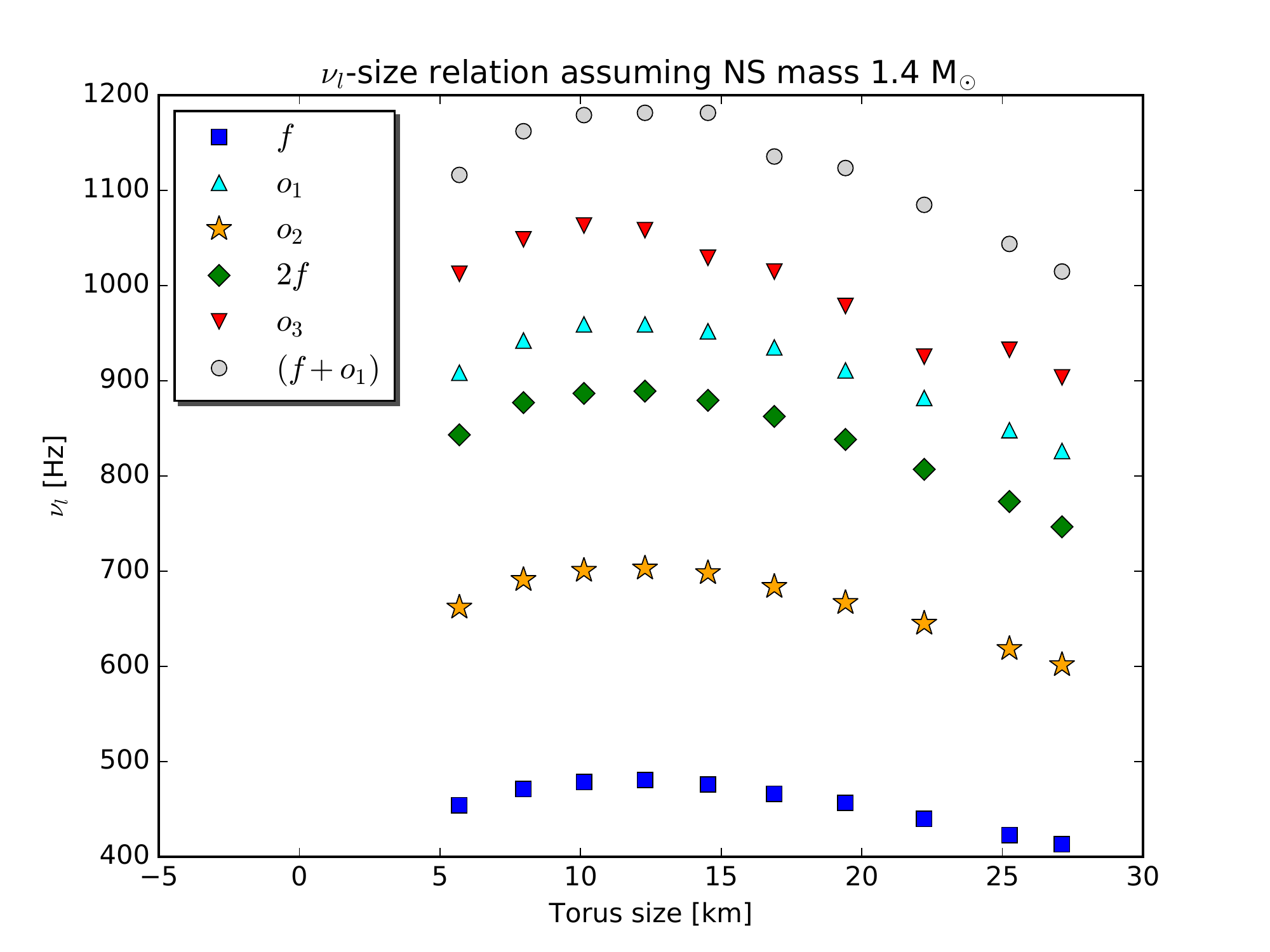}
\caption{Frequency-size relation for a neutron star of mass $M =
  1.4\,M_{\odot}$ and where different colours refer to different
  oscillation modes. Note that there is a general nonlinear behaviour
  yielding larger frequencies for smaller tori.}
\label{fig:nu_size}
\end{figure}

The chosen frequency pairs reported in
Fig. \ref{fig:qpos_as_axisymmetric_oscillations} are meant to match the
lower ($\nu_{l}$) and the upper ($\nu_{u}$) kHz QPOs. Red dots represent
the observed data while the points with colours different from red
represent the frequencies measured from our simulated tori. Clearly, all
of the simulated data sets shows a remarkably linear relation between the
chosen frequencies. In addition, we have also found that there is a
nonlinear relation between the torus size and oscillation frequency with
an overall deviation of $\sim 10\%$ over all considered models. As shown
in Fig. \ref{fig:nu_size}, larger tori display in general smaller
frequencies. However, below a critical torus size of $\simeq 12\rm km$
(for a stellar mass of $1.4 M_\odot$) all frequencies start to decrease
as the size of the torus is decreased.

Since the oscillation frequencies are related to the typical timescales
associated with the central object, and thanks to the simplicity of the
Schwarzschild metric which we use for the exterior of the relativistic
star, the only degree of freedom is the stellar mass $M$. As a result,
all of the reported frequencies can be scaled simply as $\nu_{1} /
\nu_{2} = M_{2} / M_{1}$. Stated differently, decreasing the mass of the
neutron star shifts the peaks to higher frequencies. However, although
the mass of the compact object does not change the ratio between the
peaks, i.e., $\nu_{u}/\nu_{l}$, it affects the peak separation between
these two peaks in the same ratio $M_{2}/M_{1}$. This is behaviour is
clearly shown by the dual representation in the two panels of
Fig. \ref{fig:qpos_as_axisymmetric_oscillations}, where the left panel
shows the effect of assuming a neutron-star mass of $M =
1.7\,M_{\odot}$. The right panel, on the other hand, reports the same
data and the corresponding correlations when the mass of the star is
assumed to be $M = 1.4 M_{\odot}$. In this case, one can see that three
choices of pairs of peaks provide all rather good representations of
observational relation: peaks $o_{1}$ and $2f$ (black upside down
triangles), peaks $o_{1}$ and $o_{3}$ (green diamonds) and peaks $o_{2}$
and $f+o_{1}$ (blue squares). At the same time, we should recall that
this value of the mass for the neutron star is smaller of what estimated
by \citet{Casares2006}, who have constrained the mass of the central
compact object in 4U 1636--53 to be in the range $1.6 M_{\odot} < M < 1.9
M_{\odot}$.

In addition, as can be seen in the two panels of
Fig. \ref{fig:qpos_as_axisymmetric_oscillations}, not all choices of
pairs of peaks can match the observed linear relation between the
observed frequencies. In fact, assuming the mass of the neutron star to
be in the range given by \cite{Casares2006}, \ie $M = 1.7\,M_{\odot}$, no
choice can match the observational relation except the pair of peaks
$o_{1}$ and $o_{3}$ (green diamonds) and $o_{2}$ and $f+o_{1}$ (blue
squares), but only when the neutron star is assumed to have a mass $M=
1.4\,M_{\odot}$; even in these cases, however, the match is
marginal. More specifically, in the case of peaks $o_{1}$ and $o_{3}$,
although the slope of the correlation and the peak separation seem to be
in agreement with the observational relation, the smallest physically
meaningful torus, which displays the highest frequencies in this
sequence, only reaches the lowest frequencies of the observational
relation (see left panel of Fig.
\ref{fig:qpos_as_axisymmetric_oscillations}). Similarly, the modes
$o_{2}$ and $f+o_{1}$, although promising since they match the observed
relation for high frequencies, \ie for
$[\nu_{o_{2}},\nu_{f+o_{1}}]>[740,1010]$ Hz, do not extend to the
low-frequency end of the observational relation.

\begin{figure}
\centering
\includegraphics[width=0.95 \columnwidth, angle=0]{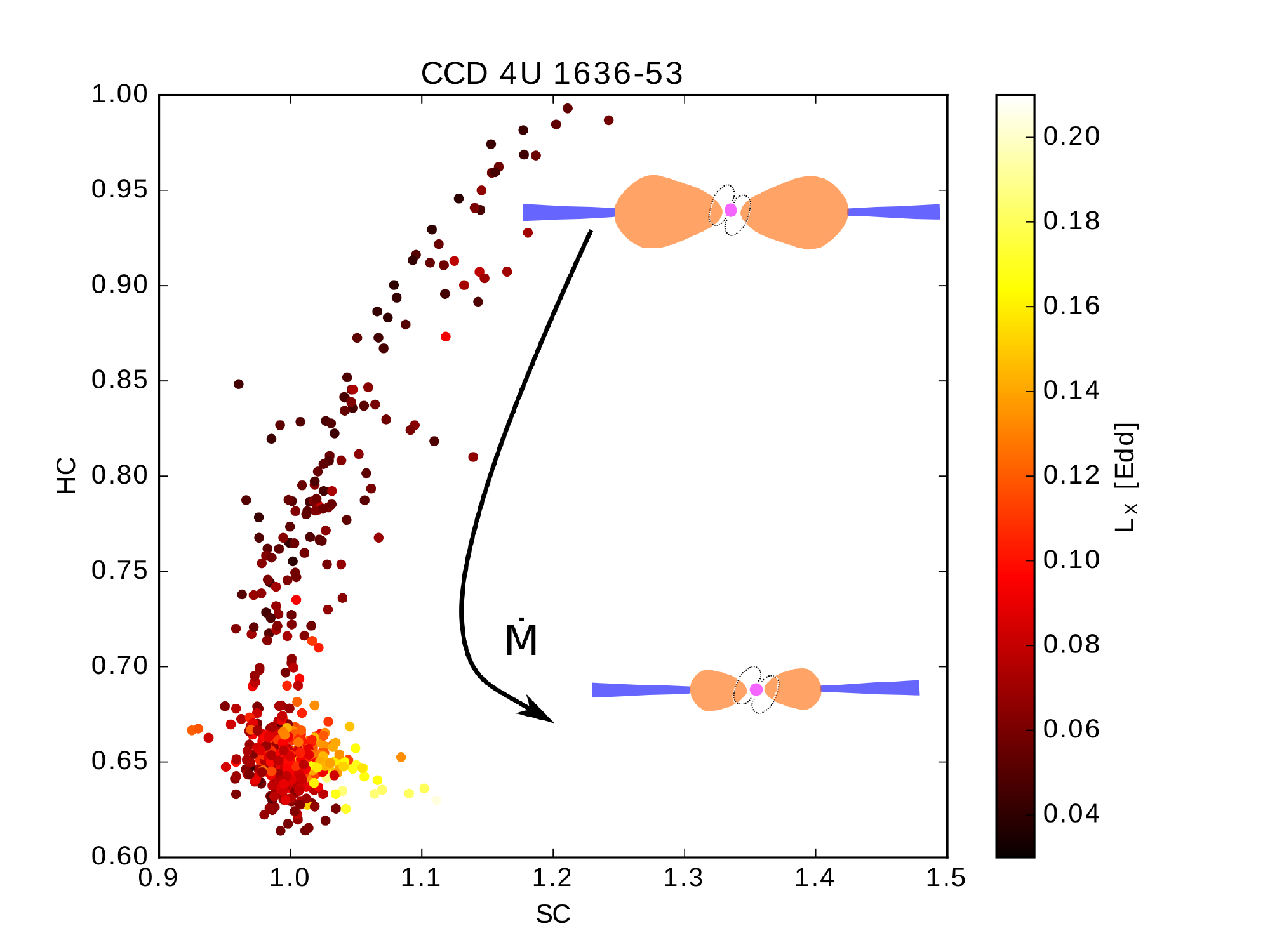}
\caption{Colour-colour diagram (CCD), where the horizontal and vertical
  axes represent the ``hard colour'', $\mathrm{HC}$ and the ``soft
  colour'', $\mathrm{SC}$, respectively. (see main text for details)
  Larger tori show lower frequencies in general and should be placed in
  the upper parts of the CCD, where the frequencies are, in general,
  lower, especially for the upper kHz QPO. Also shown schematically is
  the direction of increase of the mass-accretion rate and a cartoon
  showing two suggested states of the accretion disc: in the upper part
  the mass-accretion rate is low and and disc is farther away from the
  central object, allowing a bigger torus and lower QPOs. Conversely, in
  the lower part, the accretion rate is high and the disc is closer to
  the central object, compressing the torus and yielding higher
  QPOs. This framework can be used to explain both the observations and
  the results of the simulations (see main text for details).}
\label{fig:ccd-torus}
\end{figure}

Before concluding this section a few considerations are worth
making. First, since the constant-$\ell$ torus size is delimited from
above by the requirement that it does not penetrate the neutron star
surface (thus delimiting the frequencies from below), and the oscillation
frequencies achieve their maximum at finite torus extent, the range of
achievable frequencies is fixed by the model.

Second, we recall that we have here used the SLy4 equation of state, for
which a nonrotating star with mass $M = 1.7\,M_{\odot}$ has a radius of
$11.42$ km. On the other hand, if the star is assumed to have a mass $M =
1.4\,M_{\odot}$ the radius is $11.67$ km, thus extending the range of
sizes of the possible tori. Hence the equation of state chosen
for the neutron star does play a role in the sense that it defines,
through the radius of the star, the boundary of our simulations as well
as the maximum size of the torus. Indeed, regarding the $1.4\,M_{\odot}$
case, it would require building a torus bigger than the ones considered
so far.  Opposite considerations apply for the mode peaks $o_{1}$ and
$2f$ (black inverted triangles) and $o_{1}$ and $o_{3}$ (green
diamonds). Both pairs match the low-frequency tail of the observed linear
correlation when $M = 1.4\,M_{\odot}$, but not the high-frequency
end. Going to smaller neutron-star masses would not help much in this
case, since to increase the frequencies one would need to reduce the
torus' size by reducing the value of the specific angular momentum,
which is however very close to the minimum attainable \ie $\ell \simeq
3.67$ \citep{Rezzolla_book:2013}; in fact, as remarked above, models
smaller than \texttt{Tor.03} would be difficult to justify physically
because of their small size (\cf Table \ref{tab:tori_pars}).

%% Third, although the oscillation modes found do no match the entire
%% observed kHz QPOs frequency relation, this approximation provides more
%% information than was previously thought. This is so in part because the
%% simulated $\nu_{l}-\nu_{u}$ relation is not too dissimilar from the
%% observational relation. The observational relation arises from the
%% rapidly changing environment around the neutron star in the system where
%% both the lower and the upper kHz QPOs arise. As discussed by

Third, although the oscillation modes found do not match exactly the
entire observed kHz QPOs frequency relation, this approximation provides
more information than was previously expected. We recall that the
observed $\nu_{l}-\nu_{u}$ relation, which is detectable when both the
lower and the upper kHz QPOs appear, is thought to arise from the rapidly
changing environment around the neutron star. Indeed, the appearance of
both kHz QPOs occurs only in the left lower corner of the CCD, which is
also the region of the CCD where the system changes considerably
\citep{Zhang2017}. We show in Fig. \ref{fig:ccd-torus} the CCD for 4U
1636--36, together with the corresponding X-ray luminosity as indicated
by the colorcode in units of the Eddington luminosity.  Recalling that
the CCD depends on the definition the hard and soft colours and following
\cite{Altamirano2008}, we define the ``hard colour'' ($\mathrm{HC}$) as
the $9.7-16.0\,{\rm keV}/6.0-9.7\,{\rm keV}$ count-rate ratio and the
``soft colour'' ($\mathrm{SC}$) as the $3.5-6.0\,{\rm keV}/2.0-3.5\,{\rm
  keV}$ count-rate ratio, respectively. Note that the X-ray luminosity is
higher in the region of the CCD where both kHz QPOs are simultaneously
detected, thus enabling the detection $\nu_{l}-\nu_{u}$ observational
relation [see \citet{Zhang2017} and references therein for more
  details]. Since the $p$-mode frequencies of the tori are in first
approximation related to their size, the sequence of tori considered here
could represent different states of 4U 1636--53. More specifically, it is
plausible to imagine that an external mechanism, possibly within the
accretion flow, could determine the size of the torus at a given moment
as the source moves across the CCD, perturbing the torus in the process.

Since there is an anticorrelation between the frequencies of the QPOs and
the hard colour for Atoll sources \citep{Linares2009}, specially for the
upper kHz QPOs\footnote{In 4U 1636--53, the upper QPO varies from $\sim
  400$ Hz at $\mathrm{HC} = 1.0$ to $\sim 1220$ Hz at $\mathrm{HC} =
  0.65$; the lower QPO is detected at $\sim 550$ Hz for $(\mathrm{HC},
  \mathrm{SC})=(0.7, 0.98)$ and at $\sim 890$ Hz at $(\mathrm{HC},
  \mathrm{SC})=(0.62, 1.1)$.}, it is then tempting to place the tori in
the CCD according to their sizes and we envisage
that larger tori are placed in the upper right corner of the CCD (when
the source is in the hard state) and smaller tori are placed in the lower
left corner of the CCD (when the source is in the soft state) as depicted
schematically in Fig. \ref{fig:ccd-torus}.
%
%% It is then tempting to place the
%% tori in the CCD according to their sizes and envisage that larger tori
%% are placed in the upper right corner of the CCD (when the source is in
%% the hard state) and smaller tori are placed in the lower left corner of
%% the CCD (when the source is in the soft state) as depicted schematically
%% in Fig. \ref{fig:ccd-torus}. \lr{I find that the following text in red is
%%   not needed in any way for our discussion; it just provides numbers that
%%   anyway we don't use and confuses our discussion; I suggest we simply
%%   remove it} \textcolor{red}{In the CCD,
%%   the position of the source marks, besides other properties, the
%%   frequencies of the QPOs, with the the upper is detected througout the
%%   CCD (with changing strengths) at $\sim 400$ Hz for $\mathrm{HC} = 1.0$
%%   and at $\sim 1220$ Hz at $\mathrm{HC} = 0.65$; the lower QPO, on the
%%   other hand, is detected at $\sim 550$ Hz for $(\mathrm{HC},
%%   \mathrm{SC})=(0.7, 0.98)$ and at $\sim 890$ Hz at $(\mathrm{HC},
%%   \mathrm{SC})=(0.62, 1.1)$ [note that the lower QPO is detected only
%%     when the $\mathrm{HC}$ is $0.7$ or lower; see, for example,
%%     \cite{deAvellar2016,Zhang2017}].} 
%
We conjecture, therefore, that the variation of the X-ray luminosity
across the CCD is responsible, through the accretion rate, for changes in
size of the torus around the neutron star. This idea is reported
schematically using the data of 4U 1636--53 also in
Fig. \ref{fig:ccd-torus}: as the accretion rate increases from hard to
soft states \citep[see, for example,][and references
  therein]{Liu2005,Done2007}, so does the X-ray luminosity and the
accretion disc extends closer to the surface of the neutron
star. Consequently, the overall observational picture seems to be
consistent with placing large tori in hard states -- when the disc is
expected to be more distant and the tori can ``expand'' -- and small
tori in soft states -- when the disc is ``compressed'' as a result of
the increased accretion rate.

Finally, we can compare and contrast our results with those obtained
recently by other groups. In particular, we note that our results are in
agreement with those of \cite{Torok2016}, where the authors employ a
similar idea, but consider non-axisymmetric frequencies. Their work is
based on the analysis by \cite{Straub2009}, who derived
pressure-corrected fully general-relativistic expressions for the
eigenfunctions and eigenfrequencies of the radial and vertical epicyclic
modes of a slightly non-slender, constant specific angular momentum
torus. \cite{Torok2016} built a sequence of tori with different
thicknesses and for each model they found the modes of oscillation,
identifying the lower and upper kHz QPOs with the frequency of the
non-axisymmetric $m=-1$ radial epicyclic mode and the Keplerian frequency
at the centre of the torus, respectively. In this way they were able to
match the $\nu_{l}-\nu_{u}$ relation for the kHz QPOs considerably better
than previous models, such as the relativistic-precession model
\citep{Stella1999}, constraining the mass of the neutron star in 4U
1636--53 to be $M=1.69\,M_{\odot}$. More recently,
\cite{Parthasarathy2017} performed axisymmetric ideal-MHD Newtonian
simulations of oscillating cusp-filling tori orbiting a nonrotating
neutron star. They followed the response of the mass-accretion rate to
the oscillations of the torus and related this to the boundary layer
formed at the surface of the neutron star, finding that the most
prominent mode of oscillation in the mass-accretion rate is the radial
epicyclic mode, which they associated with the lower kHz QPO; they could
not detect, however, the upper kHz QPO. We note therefore the analogy
with our suggestion of searching an imprint of the torus' oscillations on
the mass-accretion rate, which is ultimately related to the
boundary-layer luminosity, as demonstrated by \cite{Gilfanov2003} and
\cite{Gilfanov2005}.

\section{Conclusions}
\label{discuss}

We have performed general-relativistic simulations of non-selfgravitating
and axisymmetric tori with constant specific angular momentum orbiting
around a nonrotating neutron star. After triggering oscillations via a
small radial and vertical velocity perturbation, we have followed their
evolution and setup a precise scheme for the robust identification of the
modes in the PSDs. In agreement with the linear analysis by
\citet{Rezzolla_qpo_03a,Zanotti03} and \cite{Zanotti05}, we have
identified the fundamental $p$-mode frequency and its overtones, thus
determining eight different frequencies of increasing value in the PSD of
various hydrodynamical quantities, such as the rest-mass density. The
peaks correspond to the modes of oscillation: $f$, $o_{1}$, $o_{2}$,
$2f$, $o_{3}$, $f+o_{1}$, $f+o_{2}$, and $3f$, as marked in the
classification of \citet{Zanotti05}. Also in agreement with previous work
on tori around black holes, we have found that the value of the
fundamental frequency (and hence of the overtones) is inversely
proportional to the torus size, scaling linearly with the mass of the
star. The dependence of the frequencies on the size of the torus for a
given neutron-star mass was then used to compare with the observational
relation between the upper and the lower kHz QPOs in NS-LMXB 4U
1636--53. This was done by appropriately selecting pairs of peaks in the
PSDs and relating each pair to the lower and upper kHz QPOs, $\nu_{l},
\nu_{u}$, with the goal of reproducing the observed linear relation.

Using the ten simulated models and taking into account all possible mode
pairs, we have found that as long as we restrict our analysis to tori
with constant specific-angular momentum, it is not possible to match the
entire observational $\nu_{l}-\nu_{u}$ relation, although there are pairs
of modes which either reproduce well the observed relation, \eg modes
$o_{2}$ and $f+o_{1}$, but in a limited frequency range, or have correct
linear slope, \eg modes $o_{1}$ and $o_{3}$, but that require stellar
masses $M \simeq 1.4\,M_{\odot}$ and hence smaller than what expected
from other analyses. Overall, while our simulations provide promising
evidence that the phenomenology observed in 4U 1636--53 could be
explained in terms of the $p$-mode oscillations of tori around the
neutron star, it is also clear that the torus models considered here with
a constant specific angular momentum can only provide a marginal match
with the observed data. This is essentially due to the limitations set by
the interplay between the torus size and the radius of the star.

These considerations promote at least three different directions in which
our analysis can be improved and that we will consider in future
work. First, we can extend this analysis to more general distributions of
the specific angular momentum and determine whether the different tori
sizes that will be allowed in this way will also provide a better match
to the observations. Second, we can reconsider our models using fully
three-dimensional simulations that would excite non-axisymmetric
modes. It is then possible that, with a larger number of modes available,
it will be easier to obtain a more accurate match with the observational
data. Finally, we can deepen our astrophysical understanding of LMXBs by
modelling the radiative processes thought to occur in these systems and
their subsequent emission properties and signatures. In particular, using
the results of the simulations coupled with the solution of the
general-relativistic radiative transfer problem, we can produce synthetic
spectra and light curves where time and phase lags will be an integral
part of our modelling. In this way we will be able to study other
properties of the QPOs, \eg the fractional amplitude, and correlations
with energy and frequency of the QPOs \citep[see][]{deAvellar2016}.

\section*{Acknowledgements}

MGBA acknowledges the financial support from FAPESP 2015/20553-0 and from
the FAPESP Thematic Project 2013/26258-4. Additional support comes from
``NewCompStar'', COST Action MP1304, the LOEWE-Program in HIC for FAIR,
the European Union's Horizon 2020 Research and Innovation Programme
(Grant 671698) (call FETHPC-1-2014, project ExaHyPE) and the ERC synergy
grant "BlackHoleCam: Imaging the Event Horizon of Black Holes" (Grant
No. 610058).  MGBA is also
grateful to the Institut f\"ur Theoretische Physik for the kind
hospitality and to E. Ribeiro and P. Bult for the useful comments. ZY
acknowledges support from an Alexander von Humboldt Fellowship.

%%%%%%%%%%%%%%%%%%%%%%%%%%%%%%%%%%%%%%%%%%%%%%%%%%



\bibliographystyle{mnras}
\bibliography{aeireferences} % if your bibtex file is called example.bib
%\bibliography{aeireferences,local} % if your bibtex file is called example.bib

%%%%%%%%%%%%%%%%% APPENDICES %%%%%%%%%%%%%%%%%%%%%

%\appendix

%%%%%%%%%%%%%%%%%%%%%%%%%%%%%%%%%%%%%%%%%%%%%%%%%%

% Don't change these lines
\bsp	% typesetting comment
\label{lastpage}
\end{document}